\documentstyle[12pt,amsmath,amssymb,epsf,epic]{article}

\numberwithin{equation}{section}

\newcommand{\ra}{\rightarrow}
\newcommand{\tr}{\mbox{Tr}}

\newcommand{\bra}{\langle} 

\newcommand{\ket}{\rangle}

\newcommand{\E}{{\mathbb E}}
\newcommand{\D}{{\mathbb D}}

\newcommand{\U}{{\mathbb U}}
\newcommand{\R}{{\mathbb R}}
\newcommand{\N}{{\mathbb N}}

\newcommand{\C}{{\mathbb C}}
\newcommand{\Z}{{\mathbb Z}}
\renewcommand{\E}{{\mathbb E}}
\renewcommand{\P}{{\mathbb P}}
\newcommand{\I}{{\mathbb I}}
\newcommand{\T}{{\mathbb T}}
\newcommand{\s}{{\mathbb S}}

\newcommand{\hil}{{\mathcal H}}

\newcommand{\be}{\begin{equation}}
\newcommand{\ee}{\end{equation}}
\newcommand{\bea}{\begin{eqnarray}}
\newcommand{\eea}{\end{eqnarray}}

\newcommand{\eps}{\epsilon}
\newcommand{\ffi}{\varphi}

\newcommand{\ep}{\hfill  {\vrule height 10pt width 8pt depth 0pt}}

\newcommand{\grintl}{[\kern-.18em [}
\newcommand{\grintr}{]\kern-.18em ]}

\newcounter{resultcounter}[section]

\newtheorem{theorem}[resultcounter]{Theorem}
\newtheorem{lemma}[resultcounter]{Lemma}
\newtheorem{proposition}[resultcounter]{Proposition}
\newtheorem{corollary}[resultcounter]{Corollary}

\newtheorem{remark}[resultcounter]{Remark}

\begin{document}

\title{Dynamical Localization for $d$-Dimensional Random Quantum Walks
\thanks{Partially supported by the Agence Nationale de la Recherche, grant ANR-09-BLAN-0098-01}
}

\author{ Alain Joye \footnote{
              UJF-Grenoble 1, CNRS Institut Fourier UMR 5582, Grenoble, 38402, France} 
              }

\date{  }

\maketitle

\begin{abstract}
We consider a $d$-dimensional random quantum walk with site-dependent random coin operators. The corresponding transition coefficients are characterized by deterministic amplitudes times independent identically distributed site-dependent random phases. When the deterministic transition amplitudes are close enough to those of a quantum walk which forbids propagation, we prove that dynamical localization holds for almost all random phases. This instance of Anderson localization implies that all quantum mechanical moments of the position operator are uniformly bounded in time and that spectral localization holds, almost surely.
\end{abstract}

\section{Introduction}
\label{intro}
Quantum walks defined on a lattice, deterministic or random, have become a popular topic of research recently, due the interest they have for different scientific communities, see the reviews \cite{Ke}, \cite{Ko}, \cite{AVWW}. Quantum walks can be used to provide an effective description of the dynamics of certain types of physical quantum systems and belong, in this sense, to the broader category of quantum network models, \cite{ADZ}, \cite{M}, \cite{CC}, \cite{BB}. In particular, quantum walks accurately describe the dynamics of atoms trapped in certain time dependent optical lattices, as was recently demonstrated experimentally, \cite{Ketal}, \cite{Zetal}. 
Quantum walks also offer a field of investigation for probabilists since the quantum interpretation of the wave function together with the relative simplicity of the dynamics of quantum walkers give rise to generalizations of classical random walks possessing sometimes unfamiliar features, like ballistic transport properties, see {\it e.g.} \cite{Ko}, 
\cite{APSS}.  Moreover, in the same way as classical random walks are important for classical computer science, quantum walks play an important role in the benchmarking procedures used to assess quantum algorithms in theoretical quantum computing \cite{S}, \cite{MNRS}. 

It is of interest in some of the instances mentioned above to introduce static randomness in the dynamics of the quantum walk, in order to take into account fluctuations in the model, be they unavoidable experimentally or inherent to the model,  \cite{RHK}, \cite{YKE}. In such cases, one expects the transport of the system to be completely suppressed due to the destructive interferences induced by the randomness and one speaks of localization. This well known phenomenon in solid state physics goes under the name Anderson localization and it is mathematically well established for a large variety of continuous time dynamics driven by random Schr\"odinger operators, in various regimes, see e.g. \cite{St}, \cite{Ki}. While of a similar nature, this phenomenon has been less studied for the case of discrete quantum dynamics defined by random unitary operators. Localization results have been proven for certain types of random unitary operator in \cite{BHJ} \cite{J}, \cite{Si}, \cite{HJS1}, \cite{HJS2}, \cite{ABJ1}, \cite{ABJ2} with various physical and mathematical motivations, see \cite{J2} for a review. Concerning coined random quantum walks specifically, the following contributions address such questions, \cite{Ko1}, \cite{JM}, \cite{SK}, \cite{ASW}. The first paper yields a special instance where the randomness fails to induce localization, while the last three papers provide localization results. These results hold for the dimension one only.

Our purpose in this paper is to prove dynamical localization for $d$-dimensional random quantum walks, $d\geq 1$, characterized by random transition coefficients and deterministic transition probabilities, in the same spirit as the one-dimensional model addressed in \cite{JM}. 
We make use of ground work on dynamical localization by means of fractional moments estimates for random unitary operators provided in \cite{HJS2} and of the methods used in the recent paper \cite{ABJ2} to tackle two dimensional problems in a perturbative regime analogous to the large disorder regime of the Anderson model.

Let us close this introduction by briefly mentioning two unrelated instances where one speaks of localization and of randomization in the study of quantum walks and which are not discussed in this paper. First,
under some circumstances, deterministic quantum walks that are either homogeneous in space or spatially locally perturbed by the presence of defects give rise to dynamics with the following features: An initial states is split into a part that travels to infinity ballistically and a part that remains for all times in a neighborhood of the origin. The persistence of a part of the quantum walker close to the origin, sometimes also called localization, is of a completely different nature from Anderson localization and is due to the appearance of eigenvalues in the spectrum of a relevant unitary operator, in addition to the usual continuous spectrum, see \cite{IKK}, \cite{BHJ}, \cite{CGMV}, \cite{KLS}.
The second point which is not discussed here concerns a different way of randomizing a coined quantum walk. It consists in making the coin matrices depend on time in a random way, which leads to the study of random dynamical systems for the corresponding random quantum walk. Various models of this sort have been considered in the physics literature and numerical as well as rigorous studies have been lead, see {\it e.g.} \cite{SBBH}, \cite{KBH}, \cite{AVWW}, \cite{J3}, \cite{HJ}, and references therein. The main outcome is the appearance of diffusion phenomena in the averaged dynamics, restoring a flavor of classical walk.

\section{The Model and Main Result}

We consider random quantum walks on the regular lattice $\Z^d$ with coin space $\C^{2d}$, {\it i.e.} a unitary random operator on $\hil=\C^{2d} \otimes l^2(\Z^d)\simeq l^2(\Z^d;\C^{2d})$, by means of the following construction. 

\medskip

Let us denote the standard basis of $\C^{2d}$ by 
\be\label{ipm}
\{|\tau\ket\}_{\tau\in I_\pm}, \ \ \mbox{ where} \ \ I_\pm=\{+1,-1,+2,-2, \dots, d, -d\}
\ee
and that of $l^2(\Z^d)$ by $\{|y\ket\}_{y\in \Z^d}$. The corresponding basis vectors of the tensor product $\hil$ will be written as $|\tau,y\ket=|\tau\ket \otimes |y\ket$ and orthogonal projectors on these basis vectors are denoted by $P_\tau=|\tau\ket\bra\tau|$ etc... The scalar product of $\ffi, \psi\in \hil $ is denoted by $\bra\ffi, \psi\ket$, with linearity on the right hand side, and the associated norm is denoted by $\|\ffi\|^2=\bra\ffi, \ffi\ket$. We also denote by $\|\cdot\|$ the norm of bounded operator on $\hil $. We set $r: I_\pm\ra\Z^d$ by
\be
r(\tau)=\mbox{sign}(\tau) e_{|\tau |},  \ \ \  \tau\in I_\pm, 
\ee
{with $\{e_j\}_{j=1,2,\dots, d}$ the canonical basis of $\R^d$.}

Let $S$ be the coin state dependent shift operator defined on $\hil$ by 
\be\label{ss}
S=\sum_{y\in\Z^d, \tau\in I_\pm}P_\tau\otimes |y +r(\tau)\ket\bra y|.
\ee 
  In other words, $S |\sigma, x\ket=|\sigma, x+r(\sigma)\ket$, for any basis vector $|\sigma, x\ket$.  Let $C\in U(2d)$, $U(2d)$ the set of unitary matrices on $\C^{2d}$, be a coin matrix. Then  the unitary operator $\s$ on $\hil$ given by
\be\label{dqw}
\s=S(C\otimes \I)=\sum_{y\in\Z^d, \tau\in I_\pm}P_\tau C\otimes |y + r(\tau) \ket\bra y|
\ee
defines a deterministic  quantum walk. 

Consider now $\T^{\Z^d\times I_\pm}$ as a probability space with $\sigma$ algebra generated by the cylinder sets and measure $\P=\otimes_{y\in\Z^d\atop \tau\in I_\pm}d\mu$ where $d\mu$ is a probability measure on $\T$. 
We introduce accordingly the set of i.i.d random variables $\omega^{\tau}_{y}\in \T$ labelled by ${y\in\Z^d}$ and ${\tau\in I_\pm}$, with common distribution $d\mu$.  We construct a diagonal random unitary operator
$\D(\omega)$ on $\hil$
by 
\be\label{dd}
\D(\omega)\, |\tau, y\ket =e^{i \omega^{\tau}_{y}} |\tau, y\ket.
\ee

Eventually, we define the one time step random evolution operator \\
$U_\omega(C): \hil\ra\hil$ by
 \be\label{uu}
 U_\omega(C)=\D(\omega)\s,
 \ee 
 emphasizing the dependence on the coin matrix $C\in U(2d)$ in the notation, which should be thought of as a parameter in the problem. This definition, which is the generalization of the one considered in \cite{JM} to dimensions larger than one, amounts to make the coin matrix that the walker encounters site dependent in a random fashion, so that the transition amplitudes are multiplied by a random phase. Explicitely, 
 \be\label{sdc}
 U_\omega(C)|\sigma,x\ket=\sum_{\tau\in I_\pm}e^{i \omega^{\tau}_{x+r(\tau)}}C_{\tau,\sigma}|\tau, x+r(\tau)\ket. 
 \ee 
Note that $U_\omega(C)$ defines an ergodic family of unitary operators: For any $a\in\Z^d$, let $T_a: \T^{\Z^d\times I_\pm}\ra \T^{\Z^d\times I_\pm}$ defined by $(T_a \omega)_y=\omega_{y+a}$, where $\omega_y=(\omega_y^\tau)_{\tau\in I_\pm}$. Then
\be\label{erg}
U_{T_a \omega}(C)=V_a U_\omega(C) V_{-a},
\ee
where the unitary operator $V_a: \hil\ra\hil $ is defined by $V_a |\tau, y\ket=|\tau,y+a\ket$, for all $\tau\in I_\pm$. In particular, $V_a\s V_{-a}=\s$, for all $a\in\Z^d$.\\

The proof of our results below requires some regularity of the measure $\mu$ and we assume the following comfortable hypothesis

\medskip
\noindent{\bf Assumption R:}\ The i.i.d random variables $\{\omega_y^\tau\}_{y\in\Z^d,\tau\in I_\pm}$ have an $L^\infty$  density:
\be
d\mu(\theta)=l(\theta)d\theta, \ \  \mbox{with } \ \ 0\leq l \in L^\infty(\T).
\ee

\medskip

We shall work in a "large disorder regime", meaning that the coin matrix $C\in U(2d)$ of the random walk, viewed as a parameter of the walk, is close to some $C_\pi\in U(2d)$, such that the corresponding random quantum walk $U_\omega(C_\pi)$ is strongly localized in the following sense:  $U_\omega(C_\pi)$ admits a bloc decomposition with respect to an infinite family of deterministic finite dimensional subspaces of $\hil$, the direct sum of which coincides with $\hil$. Consequently, $U_\omega(C_\pi)$ completely forbids transitions between these subspaces, and has pure point spectrum. Moreover, in order to engage in a perturbative analysis, these invariant subspaces should allow for the definition of unitary restrictions of $U_\omega(C_\pi)$, denoted by $U_\omega^{\Lambda}(C_\pi)$,  to finite volume portions $\Lambda$ of the lattice $\Z^d\times I_\pm$,  for which a detailed spectral analysis is available. \\

As we will show below, matrices from the set of permutation matrices, 
\be\label{pm}
{\cal PM}=\{ C_\pi\in U(2d) \ | \  C_\pi |\tau\ket= |\pi(\tau)\ket, \forall  \tau\in I_\pm,  \ \mbox{with } \ \pi\in{\mathfrak S}_{2d} \ \mbox{acting on } \ I_{\pm},\}
\ee
satisfy the previous requirements. Actually, it will be enough to prove the existence of invariant subspaces for $\s_\pi=S (C_\pi \otimes \I)$. Note that any permutation $\pi\in{\mathfrak S}_{2d}$ without fixed point admits a unique decomposition in cycles with disjoint supports of the form 
\be\label{pp}
\pi=\pi_1 \pi_2 \dots \pi_n,\ \mbox{where  $\pi_j=(\tau_j, \pi(\tau_j), \pi^2(\tau_j), \dots, \pi^{m_j-1}(\tau_j))$, }
\ee 
for a set of $n\leq d$ indices $\tau_j$ appearing in sequence in the set $I_\pm$ ordered as in (\ref{ipm}) with $\tau_1=1$ and  $\sum_{j=1}^n m_j=2d$. 
\\

As we prove below, our conditions on  $\s_\pi$  imply conditions on  $\pi\in{\mathfrak S}_{2d}$:
\begin{lemma}\label{nfp} The operator $\s_\pi$ has pure point spectrum if and only if $\pi\in{\mathfrak S}_{2d}$ satisfies \be\label{sumcond}
\sum_{s=0}^{m_j -1}r(\pi_j^s(\tau_j))=0, \ \ \ \forall j\in\{1,\dots, n\}.
\ee
\end{lemma}
\begin{remark}\label{fprem} In particular, $\pi$ has no fixed point. Moreover, $m_j$ is even.
\end{remark}
We can now state our main result:

\begin{theorem}\label{dynloc}
Let $U_\omega(C)=\D(\omega)\s$ be defined on $\hil =\C^{2d} \otimes l^2(\Z^d)$ by $\s=S\,(C \otimes \I)$ given by (\ref{dqw}) and $\D(\omega)$ by (\ref{dd}), under Assumptions R. Let $C_\pi\in {\cal PM}$ associated with $\pi\in {\mathfrak S}_{2d}$ satisfying (\ref{sumcond}). 

There exists a $\delta>0$ such that  for all $U_{\omega}(C)$ with $C\in U(2d)$ satisfying $\|C-C_\pi\|_{\C^{2d}}\leq \delta$ we have
\begin{enumerate}
\item $U_{\omega}(C)$ has pure point spectrum almost surely;
\item for any nonnegative $p$ and any $\psi\in \hil$ of compact support, it holds for the multiplication operator $\vert X\vert^{p}|\tau, x\ket=\vert x \vert^{p}|\tau, x\ket=\left(\sum_{i=1}^{d}\max{|x_{i}|}\right)^{p}|\tau, x\ket$ almost surely:
\[\sup_{n\in\Z}\left\Vert \vert X\vert^{p}U^{n}_{\omega}(C)\psi\right\Vert_\hil<\infty;\]
\item there exist finite constants $g>0, c>0$ such that for all $x, y\in\Z^{d}$, $\tau, \sigma\in I_\pm$
\[\E_\omega\left\lbrack\sup_{f\in C(\U), \Vert f\Vert_{\infty}\le1}\left\vert\langle \tau, x|, \, f\left(U_{\omega}(C)\right) \, |\sigma, y\rangle\right\vert\right\rbrack\le c e^{-g\vert x-y\vert},\]
where $C(\U)$ is the set of complex valued continuous functions on the unit circle $\U$.
\end{enumerate}
\end{theorem}

\begin{remark} Dynamical localization (3) implies  exponential localization (2) and spectral localization (1), see \cite{HJS2} for details.
\end{remark}
\begin{remark} Exponential localization (2) says that the state remains localized in a neighborhood of the origin, at all times $n$, with a probability to find the walker in that neighborhood arbitrarily close to one. 
\end{remark}
\begin{remark}The results proven in this paper and the method of proof certainly apply to more general unitary walks, as those discussed e.g. in \cite{AVWW}, when they are randomized in such a way that their transition amplitudes are multiplied by i.i.d. random phases. In particular, Theorem \ref{dynloc} holds true if the elements of the permutation matrix $C_\pi$ are supplemented by deterministic phases.
\end{remark}
\begin{remark}
The assumption that $C$ be close to $C_\pi$ amounts to saying that the random quantum walks we are considering are perturbations of a (trivial) walk in which a quantum walker moves on finitely many sites only.
What makes the result for the perturbed case non trivial is the fact that we have a control on the dynamics over infinite times.    
\end{remark}

The technical statement which implies Theorem \ref{dynloc} reads
\begin{theorem}\label{ame} Under the hypotheses of Theorem \ref{dynloc}, there exist finite constants $\delta>0$, $\gamma>0$ and $c>0$ such that for all $C\in U(2d)$ with $\|C-C_\pi\|\leq \delta$, all $0<s<1/3$, all $x, y\in \Z^d$ with $|x-y|>2$, all $\tau, \sigma\in I_\pm$, all $z\in \C$ with $|z|\in (1/2,2)\setminus\{1\}$, we have
\be\label{expest}
\E_\omega(|\bra  \tau, x | , (U_\omega(C)-z)^{-1}\, |\sigma, y\ket |^s)\leq c e^{-\gamma |x-y|}.
\ee
\end{theorem} 

Indeed, it is well known that the so-called fractional moments estimate (\ref{expest}) on the matrix elements of the resolvent implies dynamical localization in the self adjoint case, as shown by Aizenman and Molchanov in \cite{AM}. The adaptation of this statement to a fairly general unitary setup is provided in \cite{HJS2}, Theorem 3.2, Propositions 3.1 and 3.2. 

Essentially, the first step of the proof of Theorem \ref{ame} consists in making use of the band structure and form of disorder of $U_\omega(C)$ to get an estimate on the squares (or second moments) of the matrix elements of the resolvent in terms of the fractional moments (\ref{expest}), up to an explicit term which diverges as $|z|\ra 1$. The second step uses a form of functional calculus for unitary operators which can be controlled in terms of the second moments of the resolvent, see Proposition 5.1  and Paragraph 5.3 of \cite{HJS2}, which leads to the third statement of Theorem \ref{dynloc}. The first and second statements are consequences of the third one. \medskip
 
In dimension larger than two, such an estimate has been obtained in some large disorder regime or at the band edges of the almost sure spectrum of the ergodic random unitary operator under study, see {e.g.} \cite{J}, \cite{HJS2}, \cite{ABJ2}. When the operator under study is off-diagonal, as it is the case for $U_\omega(C)$, the simpler method of \cite{J} cannot be applied. As shown in \cite{ABJ2} which addresses the Chalker-Coddington model, a two dimensional random unitary model which is off-diagonal, an Aizenman-Molchanov estimate can be derived in a large disorder regime from an analysis of the resolvent of finite volume restrictions of the operator. While the Chalker-Coddington model differs from the coined walks considered here, we can follow the same route.

\bigskip

The rest of the paper, devoted to the proof of Theorem \ref{ame} along the lines of  \cite{HJS2} and \cite{ABJ2}, is split in several subsections which detail the main steps of the arguments, generalizing some aspects along the way.

\section{Proofs}

By convention, we shall repeatedly use the same symbol $c$ to denote constants which may vary from line to line but depend only on unessential quantities for our purpose, as will be clear from the context. Also, we shall not distinguish anymore between symbols that denote scalar products or norms in different Hilbert spaces. The spectrum of an operator $A$ on some Hilbert space is denoted by $\sigma(A)$.

\subsection{Properties of $U_\omega(C_\pi)$}

We prove Lemma \ref{nfp}, starting with a generalization of Remark \ref{fprem}. We show that if $\s$ is defined by means of an arbitrary $C\in U(2d)$ is pure point, then $C$ is off-diagonal.  We take advantage of the translation invariance of the operator $\s$ to make use of Fourier transform arguments.\\

{\bf Proof of Lemma \ref{nfp}:} For any $\psi=\sum \psi_\tau(y)|\tau, y\ket \in\hil$, we define the Fourier transform ${\cal F} \psi$ by 
\bea
{\cal F} \psi(x)&=&\widehat\psi(x)=\sum_{\tau\in I_\pm}\widehat\psi_\tau(x) |\tau\ket\in L^2(\T^d; \C^{2d}), \ \mbox{where}\nonumber\\
\widehat\psi_\tau(x)&=&\sum_{y\in\Z^d}\psi_\tau(y)e^{-ixy}, \ x\in\T^d\subset \R^d. 
\eea
Consequently, the Fourier image of $\s$, ${\cal F}\s{\cal F}^{-1}$ on  $L^2(\T^d; \C^{2d})$, is computed as
\bea
({\cal F}\s \psi)(x)&= &{\cal F}\sum_{y, y'\in\Z^d, \tau, \tau'\in I_\pm}P_{\tau'} C\otimes |y' + r(\tau') \ket\bra y'|  \psi_\tau(y)|\tau\ket\otimes |y\ket \\
&=&\sum_{y\in\Z^d, \tau, \tau'\in I_\pm} C_{\tau', \tau}  \psi_\tau(y) e^{-ix(y + r(\tau'))} | \tau' \ket\nonumber\\
&=& \sum_{\tau'\in\I_\pm}e^{-ixr(\tau')}(C\widehat\psi(x))_{\tau'}|\tau' \ket= \sum_{\tau'\in\I_\pm}e^{-ixr(\tau')}(C({\cal F}\psi)(x))_{\tau'}|\tau' \ket.\nonumber
\eea
Hence, ${\cal F}\s{\cal F}^{-1}$
acts on  $L^2(\T^d; \C^{2d})$ as a multiplication operator by the unitary matrix 
\be
\widehat C(x)=\Phi(x)C, \ \mbox{where} \ \Phi(x)_{\sigma,\tau}=\delta_{\sigma,\tau}e^{-ixr(\tau)}\ \mbox{ is diagonal}.
\ee 
Since $x\mapsto \widehat C(x)$ is a trigonometric polynomial, ${\cal F}\s{\cal F}^{-1}$ is absolutely continuous unless the eigenvalues of $\widehat C(x)$ are all independent of $x\in\T^d$. In particular $\tr \widehat C(x)$ is independent of $x$, so that $\partial_{x_j}\tr \widehat C(x)= -i e^{-ix_j}{C}_{j,j}+ie^{ix_j}{C}_{-j,-j}\equiv 0$, $j=1,2,\cdots, d$. Hence ${C}_{j,j}={C}_{-j,-j}=0$, {\it i.e.} $C$ is off-diagonal.
Specializing  to the case $C=C_\pi$, we get the result stated in Remark \ref{fprem}. 
\medskip

We consider now $C=C_\pi$. Let $\pi_j$ be a cycle in the decomposition (\ref{pp}). Consequently, for any $j\in\{1,\dots,n\}$, any $\tau_j\in \mbox{supp }\pi_j$
\be
\s_\pi^{m_j} |\tau_j, 0\ket=\big|\tau_j,\sum_{s=0}^{m_j -1}r(\pi_j^s(\tau_j))\big\ket,
\ee
which, in Fourier space is equivalent, for any $x\in \T^d$, to the eigenvalue equation 
\be
({\cal F}\s_\pi^{m_j}{\cal F}^{-1})(x)|\tau_j\ket=e^{-ix\sum_{s=0}^{m_j -1}r(\pi_j^s(\tau_j))}|\tau_j\ket.
\ee
As above, for $\s_\pi$ to be pure point, we need $\sum_{s=0}^{m_j -1}r(\pi_j^s(\tau_j))=0$.

Conversely, set for any $x\in\Z^d$,  any $\tau_j\in \mbox{supp }\pi_j$
\be\label{hjx}
\hil^{\tau_j}_{x}=\mbox{span}\Big\{ \Big| \pi^{t}(\tau_j), x+\sum_{s=0}^{t}r(\pi^s(\tau_j))\Big\ket, t=0,1,\dots,m_j-1\Big\}. 
\ee
Since the sequence of vectors in $\hil_x^{\tau_j}$ corresponds to the successive images of $|\tau_j, x\ket$ by $\s_\pi$,
$\hil_x^{\tau_j}$ is invariant under $\s_\pi$ for any $j$ if $\sum_{s=0}^{m_j -1}r(\pi^s(\tau_j))=0$. 
Note that $\hil_x^{\tau_j}=\hil_y^{\sigma}$ iff $|\sigma, y\ket\in \hil_x^{\tau_j}$.
Let
\be
\hil^j_x=\bigoplus_{\tau_j\in\mbox{\small supp }\pi_j} \hil^{\tau_j}_x, \ \ \ \mbox{and} \ \ \hil_x=\bigoplus_{j\in\{1,\dots,n\}} \hil^j_x.
\ee
In case $\pi$ consists in one cycle only, i.e. $\pi=\pi_1$, $\hil_x=\hil_x^1$. The next Lemma, whose proof we omit and which follows from the fact that $\pi$ is a bijection, shows that $\s_\pi$ is pure point.\ep

\begin{lemma} Assuming (\ref{sumcond}),  the subspaces $\hil_x^{j}\subset \hil$, with $\dim \hil_x^{j}=m_j^2$ are such that $\hil_x^{j}\perp \hil_x^{k}$ if $j\neq k$ and $\s_\pi (\hil^{j}_x) = \hil^j_x$. Moreover, $\hil_x$ with $4n\leq \dim \hil_x\leq 4d^2$ also satisfies $\s_\pi (\hil_x) = \hil_x$ and $\hil=\bigoplus_{x\in\Z^d} \hil_x$.
\end{lemma}

\begin{remark} The last sum is not direct because $\hil_x\cap\hil_y\not=\emptyset$ in general.
\end{remark} 
\begin{remark} Since $\D(\omega)$ is diagonal, the random operator $U_\omega(C_\pi)=\D(\omega)\s_\pi$ is reduced by all subspaces $\hil_x^{\tau_j}$, $\hil^j_x$ and $\hil_x$ as well. 
\end{remark}

We shall need the following probabilistic property of the  spectrum of the reductions $U_\omega(C_\pi) |_{\hil^{\tau_j}_x}$. 
\begin{lemma} \label{reshxj} There exists $c_j<\infty$ such that
for any arc  $A\subset \U$ of small enough length, we have
\be
\P(\sigma({U_\omega}(C_\pi) |_{\hil^{\tau_j}_x})\cap A=\emptyset)\geq 1-c_j|A|.
\ee
\end{lemma}
{\bf Proof: }
The eigenvalues of $U_\omega(C_\pi) |_{\hil^{\tau_j}_x}$ can be  obtained from the property, \be
{U_\omega}(C_\pi)^{m_j} |_{\hil^{\tau_j}_x}=\det(\D(\omega) |_{\hil^{\tau_j}_x})\I |_{\hil^{\tau_j}_x}=
e^{i\theta_x^{\tau_j}(\omega)}\I |_{\hil^{\tau_j}_x},
\ee
see (\ref{pp}), where 
\be
\theta_x^{\tau_j}(\omega)=\omega_x^{\tau_j}+\omega_{x+r(\pi(\tau_j))}^{\pi(\tau_j)}+\dots +\omega_{x+r(\pi(\tau_j))+\dots r(\pi^{m_j-1}(\tau_j))}^{\pi^{m_j-1}(\tau_j)}.
\ee
Hence 
\be
\sigma({U_\omega}(C_\pi) |_{\hil^{\tau_j}_x})=e^{i\theta_x^{\tau_j}(\omega)/m_j}\{1, e^{2i\pi/m_j}, \dots, e^{2i\pi(m_j-1)/m_j}\}.
\ee
Note that $\{\theta_{x}^{\tau_j}(\omega)\}_{x\in\Z^d}^{{\tau_j}\in \mbox{\small supp }\pi_j}$ are random variables distributed according to
\be
\theta_{x}^{\tau_j}(\omega)\simeq L_j(\theta)d\theta:=(l*l*\cdots l)(\theta)d\theta, \ \mbox{$m_j$-fold convolution.}
\ee
Consequently, the eigenvalues of ${U_\omega}(C_\pi) |_{\hil^j_x}$ are correlated, and for any arc $A\subset \U$ of length $|A|<2\pi/m_j$ we compute the probability that no eigenvalue belongs to $A$ as follows:
\be
\P(\sigma({U_\omega}(C_\pi) |_{\hil^{\tau_j}_x})\cap A=\emptyset)=1-\sum_{r=0}^{m_j-1}\int_{m_j e^{-2i\pi r/m_j}A} L_j(\theta)d\theta.
\ee
In particular, Assumption R implies that there exists $c_j<\infty $ such that  $\sum_{r=0}^{m_j-1}\int_{m_j e^{-2i\pi r/m_j}A} L_j(\theta)d\theta\leq c_j |A|$, as $|A|\ra 0$.\ep\\

Considering the restriction of $U_\omega(C_\pi)$ to $\hil_x=\oplus_{j=1,\cdots,n}\hil_x^j$, we get from the independence of the $\theta_x^{\tau_j}(\omega)$'s associated with the invariant orthogonal subspaces which add up to $\hil_x$
\begin{corollary} \label{cesgap}
There exists $c<\infty$ such that
for any arc  $A\subset \U$ of small enough length, we have
\be\label{esgap}
\P(\sigma({U_\omega}(C_\pi) |_{\hil_x})\cap A=\emptyset)\geq (1-c|A|)^c.
\ee

\end{corollary}

\subsection{Finite Volume Restrictions}

Let us now turn to the definition and analysis of finite volume restrictions of $U_\omega(C)$. In order to do that, we consider a generalization of the deterministic quantum walks (\ref{dqw}) which consists in allowing the coin matrix to dependent on the site of the lattice. Let ${\cal C}=\{C_x\in U(2d)\}_{x\in\Z^d}$, and define a unitary operator on $\hil$ by 
\be
\s({\cal C})=\sum_{y\in\Z^d, \tau\in I_\pm}P_\tau C_y\otimes |y+r(\tau)\ket\bra y|.
\ee
\begin{remark} The operator $U_\omega(C)$ can be expressed as $\s({\cal C}(\omega))$ where
\be
{\cal C}(\omega)=\{C_x(\omega)\in U(2d)\}_{x\in\Z^d}, \ \mbox{with $(C_x(\omega))_{\tau,\sigma}=e^{i\omega_\tau^{x+r(\tau)}}C_{\tau,\sigma}$},
\ee
 see (\ref{sdc}).
\end{remark}

Let $\Lambda_L=\{x\in \Z^d \ | \ |x|\leq L\}$ be a box of side length $2L+1$ in $\Z^d$, where $|x|=\max_{i=1,\cdots,d}|x_i|$. Given $\pi\in {\mathfrak S}_{2d}$ without fixed point, and $L\in \N$, we set 
\bea\label{defsl}
&&\s^{L}=\s({\cal C}), \ \ \mbox{where} \ C_x=\left\{\begin{matrix}C_\pi & \mbox{if} \ |x|\in\{L-1,L,L+1\}\cr
C& \mbox{otherwise} 
\end{matrix}\right. \ \mbox{and}\\
&&U^L_\omega(C)=\D(\omega)\s^L.
\eea
The finite rank difference $U_\omega(C)-U^L_\omega(C)=T_\omega^L$ is given by
\be\label{tl}
T_\omega^L=\D(\omega)\sum_{y \in\Z^d, \tau\in I_\pm \atop  |y|\in\{L,L\pm 1\}}P_\tau (C-C_\pi)\otimes |y+r(\tau)\ket\bra y|,
\ee
so that, by the Schur condition, see \cite{K} p.143, for some constant $c>0$ uniform in the disorder,
\bea\label{pt}
\|T_\omega^L\|&\leq& c\|C-C_\pi\|. 
\eea
Let 
\be
\hil^{\Lambda_L}=\bigoplus_{\{x\; | \; |x|\leq L\}}\hil_x \ \ \mbox{and} \ \ \hil^{\Lambda_L^C} =\bigoplus_{\{x\; | \; |x|> L\}}\hil_x,
\ee
where the sums are not direct sums.
\begin{proposition}\label{invariant}
The operators $\s^{L}$ and $U^L_\omega(C)$ leave the subspaces $\hil^{\Lambda_L}$ and $\hil^{\Lambda_L^C}$ invariant.
\end{proposition}
{\bf Proof: }
It is enough to consider $\s^L$, since $\D(\omega)$ is diagonal, and to show the invariance of the finite dimensional subspace $\hil^{\Lambda_L}$, since $\s^L$ is unitary. First note that our choice (\ref{defsl}) ensures that $\bigoplus_{|x|=L}\hil_x$ is invariant under $\s^L$. Then, because of (\ref{hjx}), any vector $|\tau, y\ket$ with $|y|=L$ belongs to some $\hil_x$, with $|x|\in\{L-1, L, L+1\}$ such that $\hil_x$ is invariant under $\s^L$. Hence,  vectors $|\tau, y\ket$ with $|y|=L-1$ are mapped by $\s^L$ to vectors $|\tau', y'\ket$ with either $|y'|\in\{L-2, L-1\}$, that is to vectors of $\hil^{\Lambda_L}$  or $|y'|=L$, which belong to some invariant subspace $\hil_{x'}$, hence $|x'|\in\{L-1, L\}$.  
Consequently,  $\hil^{\Lambda_L}$ and $\hil^{\Lambda_L^C}$ are invariant.
\ep\\

We define the restriction of $U_\omega(C)$ to the finite box $\Lambda_L$ and to its complement $\Lambda_L^{C}$  as
\be\label{fv}
U^{\Lambda_L}_\omega(C)=U^L_\omega(C) |_{\hil^{\Lambda_L}}\ \ \mbox{and} \  \
U^{\Lambda_L^C}_\omega(C)=U^L_\omega(C) |_{\hil^{\Lambda_L^C}}.
\ee
\begin{remark}
The restrictions to boxes centered at arbitrary points $v$ of the lattice, $\Lambda_L+v$, have exactly the same properties as those stated below, thanks to the ergodicity of $U_\omega(C)$, see (\ref{erg}). Hence, without loss of generality, we can consider $\Lambda_L$ only.
\end{remark}

We make use of perturbation theory in $C-C_\pi$, see (\ref{pt}), to analyze the properties of the resolvent of the finite volume restriction  $U^{\Lambda_L}_\omega(C)$ by comparison with that of  $U^{\Lambda_L}_\omega(C_\pi)$. We shall use the following notations for the corresponding resolvents, for $z\not\in\U$, $C\in U(2d)$
\be
R_\omega^\#(C,z)=(U^\#_\omega-z)^{-1}, \  \#\in\{\Lambda_L, L\}.
\ee
We need a probabilistic estimate on the position of the eigenvalues of the unitary random matrix $U^{\Lambda_L}_\omega(C_\pi)$. Let $\mbox{dist}(z,B)$ denote the distance between a point $z\in \C$ and a set $B\subset \C$.

\begin{lemma}\label{grap}
For any $z\not \in \U$, any $L\in\N$, and $\eta>0$ such that $\eta L^{d}$ is small enough,
\be
\P(\mbox{\em dist}(z, \sigma(U^{\Lambda_L}_\omega(C_\pi) ) )\leq \eta)\leq c\eta L^{d}
\ee
\end{lemma}
{\bf Proof:} This lemma, based on Corollary (\ref{cesgap}), is a generalization to dimension $d$ of Proposition 2.2 in \cite{ABJ2} which deals with a two dimensional case. We only sketch the argument, omitting the details. There are $(2L+1)^d$ invariant subspaces $\hil_x$ in $\hil^{\Lambda_L}$, for which (\ref{esgap}) holds. By independence, for any  $A\subset \U$ with $|A|$ small enough, 
\be
\P(\sigma({U^{\Lambda_L}_\omega}(C_\pi))\cap A=\emptyset)\geq (1-c|A|)^{c(2L+1)^d}.
\ee
Now, for $z\not\in U$ and $\eta>0$ small, a ball centered at $z$ of radius $\eta$ intersects $\U$ on an arc of length of order $\eta$, if not empty.  Hence, 
\be
\P(\mbox{dist}(z, \sigma(U^{\Lambda_L}_\omega(C_\pi))) \leq \eta )\leq 1-
(1-c\eta)^{cL^d}\leq c\eta L^d,
\ee
when $\eta$ and $\eta L^d$ are small enough.
\ep\\

Next, we show that moments of fractional powers of the matrix elements of the finite volume resolvent $R_\omega^{\Lambda_L}(C,z)$ display arbitrary fast polynomial decay in $1/L$, for $L$-dependent small values of $\|C-C_\pi\|$.

\begin{proposition}\label{finitevolresest}
For any $s\in(0,1)$, any $a \geq 0$, any $p>1/(1-s)$ and any $C\in U(2d)$ such that $\|C-C_\pi\|\leq {1}/{L^{2(ap+d)+a/s}}$, there exists a $c>0$ such that
\be\label{fvfme}
\E_\omega\left(\left\vert\langle \tau, x|, R_{\omega}^{\Lambda_{L}}(C,z) | \sigma, y\rangle\right\vert^{s}\right)\leq\frac{c}{ L^{a}},
\ee
for all $z\not\in \U$, all $L\in\N$, all
 $| \tau, x\ket, |\sigma, y\ket \in\hil_-^{L}$ with $\vert x-y\vert>2$.
 \end{proposition}
 \begin{remark} The proof actually yields the more detailed estimate (\ref{detest}) under the conditions (\ref{cond}), which provides two terms: one depending on $R_\omega^{\Lambda_L}(C_\pi,z)$ only,  via Lemma \ref{grap}, the other containing the effect of the perturbation $C-C_\pi$.
 \end{remark}
{\bf Proof:} We adapt the strategy of \cite{ABJ2} to our $d$-dimensional model.  For $a=0$, the estimate holds for any $C\in U(2d)$, by Theorem 3.1 in \cite{HJS2}, a result based on spectral averaging which holds for general unitary operators  of the form $\D({\omega})S$ with $\D({\omega})$ diagonal as in (\ref{dd}) and $S$ deterministic, banded and shift invariant. The statement reads in our case: for all $ s\in(0,1)$ and all  $C\in U(2d)$,  there exists $ c>0$ such that  for all\ $| \tau, x\ket, |\sigma, y\ket,$ and all $ z\not\in\U$:
\begin{equation}\label{thm31}
 \E\left(\left\vert\langle \tau, x|,R^{\#}_{\omega}(C,z) |\sigma, y\rangle\right\vert^{s}\right)\leq c,
\end{equation}
where $R^{\#}$ stands either for the full resolvent $R$ or for $R^{\Lambda_{L}}$, with arbitrary $L\in\N$. For $a>0$ we use the invariance of the subspaces $\hil_x$ by $U_{\omega}(C_\pi)$, first order perturbation theory and Lemma \ref{grap}. 

We introduce the notation 
\be
|\tau, x\ket = |\alpha\ket, \ \mbox{with } \ \alpha = (\alpha_1, \alpha_2)=(\tau,x)\in \Z^{d}\times I_\pm.
\ee
Let  $\alpha\sim\beta$ denote the property $\alpha, \beta\in \Z^{d}\times I_\pm$ are in the same invariant subspace:
\be
\alpha\sim\beta\ \Longleftrightarrow\exists\ \hil_x \ \mbox{ such that } |\alpha\ket, |\beta\ket \in\hil_x.
\ee
Note that $|\tau, x\ket\sim |\sigma, y\ket$ only implies $|x-y|\leq 1$ in $\Z^d$,  and  remark that
\be
\langle {\alpha}|,R_{\omega}^{\Lambda_{L}}(C_\pi,z)|{\beta}\rangle=0 \hbox{ \rm if } \alpha\not\sim\beta. 
\ee
Then we use the resolvent identity 
\be
R_{\omega}^{\Lambda_{L}}(C,z)=R_{\omega}^{\Lambda_{L}}(C_\pi,z)+R_{\omega}^{\Lambda_{L}}(C)\left(U^{\Lambda_{L}}(C_\pi)-U^{\Lambda_{L}}(C)\right)R_{\omega}^{\Lambda_{L}}(C_\pi,z),
\ee
the fact that only neighboring states are coupled by $U_{\omega}(C)$, {\it i.e.}:
\be
\langle {\alpha}|,U^{\Lambda_{L}}_{\omega}(C)|{\beta}\rangle=0\quad \mbox{  if } \vert\alpha_2-\beta_2\vert \geq2
\ee
and the estimate
\be
\Vert U^{\Lambda_{L}}_{\omega}(C)-U^{\Lambda_{L}}_{\omega}(C_\pi)\Vert\leq c\|C-C_\pi\|,
\ee
similar to (\ref{pt}), to control the the matrix elements of $R^{\Lambda_{L}}_{\omega}(C,z)$: 
For $|\mu\ket,|\nu\ket\in\hil^{\Lambda^L}$, such that $\vert\mu_2-\nu_2\vert\ge2$, $z\not\in\U$
\bea\label{resolvent}
&&\left\vert\langle {\mu}| ,R^{\Lambda_{L}}_{\omega}(C,z) |{\nu}\rangle\right\vert\le \\
&&\sum_{\alpha,\beta\in\Z^{d}\times I_\pm \atop  \beta\sim\nu, \ \vert\alpha_2-\beta_2\vert\leq 1}\hspace{-.5cm}\left\vert\langle {\mu}|,R^{\Lambda_{L}}_{\omega}(C,z) |{\alpha}\rangle\langle {\alpha}|,\left(U_{\omega}(C)-U_{\omega}(C_\pi)\right) |{\beta}\rangle\langle {\beta}| , R^{\Lambda_{L}}_{\omega}(C_\pi,z) |{\nu}\rangle\right\vert \nonumber\\
&&\le c \frac{\|C-C_\pi\|}{{\rm dist}\left(z,\sigma\left(U_{\omega}^{\Lambda_{L}}(C_\pi)\right)\right)}\sup_{{\alpha,\beta\in\Z^{d}\times I_\pm \atop \beta\sim\nu, \  \vert\alpha_2-\beta_2\vert\leq 1}}\left\vert\langle {\mu}|,R^{\Lambda_{L}}_{\omega}(C,z)| {\alpha}\rangle\right\vert, \nonumber
\eea
where $c$ is a numerical constant as the number of sites in the above sum is finite, independent of $L$. To control the denominator, we consider for $z\not \in\U$ and $\eta>0$, the probabilistic event
\be
G_{\eta}(z):=\left\lbrace\omega\in\T^{\Z^{d}\times I_\pm}, {\rm dist}\left(z,\sigma\left(U_{\omega}^{\Lambda_{L}}(C_\pi)\right)\right)>\eta\right\rbrace
\ee
and $G^{c}_{\eta}(z)$, its complement. Recall that by Lemma \ref{grap} we have that if  $\eta L^d$ is small enough,
$\P\left(G^{c}_{\eta}(z)\right)\leq c\eta{L}^d.$
Denote by $\chi_{A}$ the characteristic function of the set $A$. 
Now, for $0<s<1$,  $p>1/(1-s)$ and  $q$ such that ${1}/{p}+{1}/{q}=1$, we have ${q}s<1$ and by H\"older's inequality and estimate (\ref{thm31}) 
\bea
&&\E_\omega\left(\chi_{G_{\eta}^{c}(z)}(\omega)\left\vert\langle {\mu}| ,R^{\Lambda_{L}}_{\omega}(C,z) |{\nu}\rangle\right\vert^{s}\right)\le\\ \nonumber
&&\P\left(G_{\eta}^{c}(z)\right)^{\frac{1}{p}}\E_\omega\left(\left\vert\langle {\mu}| ,R^{\Lambda_{L}}_{\omega}(C,z) |{\nu}\rangle\right\vert^{sq}\right)^{\frac{1}{q}}\leq c \left(\eta {L}^d\right)^{\frac{1}{p}}.
\eea

By perturbation theory, for some $c>0$,  $\|C-C_\pi\|\leq c\eta$ ensures that
\be
{\rm dist}\left(\sigma\left(U^{\Lambda_{L}}_{\omega}(C_\pi)\right),z\right)>\eta\Longrightarrow {\rm dist}\left(\sigma\left(U^{\Lambda_{L}}_{\omega}(C)\right),z\right)>\frac{\eta}{2}.
\ee
Thus, by inequality (\ref{resolvent}) 
we can estimate the complementary part
\bea
&&\chi_{G_{\eta}(z)}(\omega)\left\vert\langle {\mu}|,R^{\Lambda_{L}}_{\omega}(C,z) |{\nu}\rangle\right\vert^{s}\le\\
&&\chi_{G_{\eta}(z)}(\omega)\left\vert\frac{c\|C-C_\pi\|}{{\rm dist}\left(z,\sigma\left(U_{\omega}^{\Lambda_{L}}(C_\pi)\right)\right){\rm dist}\left(z,\sigma\left(U_{\omega}^{\Lambda_{L}}(C)\right)\right)}\right\vert^{s}\leq c\frac{\|C-C_\pi\|^s}{\eta^{2s}}, \nonumber
\eea
hence, 
\be\E_\omega\left(\chi_{G_{\eta}(z)}(\omega)\left\vert\langle {\mu}| ,R^{\Lambda_{L}}_{\omega}(C,z) |{\nu}\rangle\right\vert^{s}\right)\leq c\frac{\|C-C_\pi\|^s}{\eta^{2s}} .
\ee

Gathering these estimates, we eventually obtain for any $0<s<1$, $p>1/(1-s)$, $L\in N^*$, and any $\eta>0$ such that
\be\label{cond}
\eta L^d \ \mbox{small enough and }\ \|C-C_\pi\|\leq c\eta,
\ee
\be\label{detest}
\E_\omega\left(\left\vert\langle \tau, x|, R_{\omega}^{\Lambda_{L}}(C,z) | \sigma, y\rangle\right\vert^{s}\right)\leq c\left((\eta L^d)^{1/p}+\frac{\|C-C_\pi\|^s}{\eta^{2s}}\right).
\ee
For $a>0$ arbitrary, we first fix the scale $\eta$ by $\eta=c_0/L^{ap+d}$ for some $c_0>0$ to deal with the first term and then we fix scale of the size of the difference $C-C_\pi$ according to $\|C-C_\pi\|=1/L^{2(ap+d)+a/s}$. The constant $c_0$ can be chosen uniform in $L\in \N^*$ so that both conditions (\ref{cond}) are met, and the proof of the proposition is complete.
\ep

\bigskip
Finally, we argue that the fractional moment estimates of the finite volume resolvent (\ref{fvfme}) imply exponential decay of the fractional moment estimates of the full resolvent stated as Theorem  \ref{ame}.
This quite general property which does not depend on the peculiarities of the model is based on geometric resolvent identities which relate the full resolvent to its finite volume restriction and on decoupling/resampling arguments which yield estimates on expectations of product of fractional powers of resolvents in terms of products of expectations of fractional powers of resolvents. The self-adjoint case dealt with in \cite{AENSS} was adapted to a unitary setup which is relevant to the present case and to that considered in \cite{ABJ2}, by \cite{HJS2},  Section 13.  

We describe the main steps, relying on \cite{HJS2} for the resampling/decoupling estimates, slightly improving the argument by taking into account the dependence of the estimates in the norm of the difference between $U_\omega(C)$ and $U_\omega^L(C)$, see (\ref{tl}), which is small in our case. 

\subsection{Infinite vs. Finite Volume Resolvents}

The geometric resolvent identity relates the resolvents of $U_{\omega}(C)$, $U_{\omega}^{\Lambda_L}(C)$ and $U_{\omega}^{\Lambda_L^c}(C)$. We drop the dependences in $C\in U(2d)$, $\omega$ and $z\not\in\U$ with $1/2<|z|<2$ to simplify the notation below. Definitions (\ref{tl}) and (\ref{fv}) yield
\be
U = U^L+T^L=U^{\Lambda_L} \oplus U^{\Lambda_L^c} + T^{L},
\ee
We keep track of the dependence in $t=\|T^L\|$, where $t\leq c \|C-C_\pi\|$, uniformly in $L$ and $\omega$.
We note $R = (U-z)^{-1}$ and
\be\label{dirres}
 R^{L}= (U^{\Lambda_L} \oplus U^{\Lambda_L^c}-z)^{-1} = (U^{\Lambda_L}-z)^{-1} \oplus (U^{\Lambda_L^c}-z)^{-1}.
 \ee
 
 \begin{proposition}\label{ivsf} 
 For every $s\in (0,1/3)$ there exists a constant $c <\infty$ depending on $s$, such that
\bea \label{itstep}
&&\E(|\bra \tau, 0|,  R |\sigma, y\ket|^s) \le c t^{2s}  (1+ct^sL^{d-1} )\\ \nonumber
&&\hspace{1.cm}\times \sum_{|u\ket \in \hil \atop L-2\leq |u_2|=L+1} \E(|\bra \tau, 0|,  R^L |u\ket|^s)\sum_{|x'\ket\in\hil \atop L+2\leq |x'_2|\leq L+4} \E(|\bra x'|,  R |\sigma, y\ket|^s)
\eea
uniformly in $z\not\in\U$ with $1/2< |z| < 2$, $L\in \N$ and $y\in \Z^d$ with $|y|\ge L+5$.
\end{proposition}
\begin{remark}
In the general case where the matrix elements of $T^L$ are of order one, one can replace the prefactor by $cL^{d-1}$.  Since Proposition \ref{finitevolresest} provides arbitrary fast polynomial decay in $1/L$ of the fractional moments of the finite volume resolvent, this will turn our to be sufficient for the argument to follow, as in the case \cite{ABJ2}. Nevertheless, in the sequel, we take advantage of the fact that the matrix elements of $T^L$ are of order $t\leq c\|C-C_\pi\|$, which is small in our case, see (\ref{cond}), in order to optimize on the exponents. 
\end{remark}

 {\bf Proof :} Recall that $T^{L}$ has non-vanishing matrix-elements only near the boundary of the box $\Lambda_L$:
\bea
 T^{L} |\alpha\ket &=& 0 \quad \mbox{if $|\alpha_2|\le L-2$ or $|\alpha_2| \geq L+2$}\nonumber \\
 {T^{L}}^* |\alpha\ket &=&0 \quad \mbox{if $|\alpha_2|\le L-3$ or $|\alpha_2| \geq L+3$}
 \eea
and
\be
 \bra \beta|,  T^{L} |\alpha\ket =0 \quad \mbox{if $|\alpha_2-\beta_2| \geq  2$}.
 \ee

We do a double-decoupling, once on $\Lambda_L$ and once on $\Lambda_{L+3}$. Using the resolvent identity twice yields
\bea \label{dri}
R & = & R^{L} - R^{L} T^{L} R \nonumber \\
& = & R^{L} - R^{L} T^{L} R^{L+3} + R^{L} T^{L} R T^{L+3} R^{L+3}.
\eea
Since for $y\in \Z^d$ with $|y|\ge L+5$ and any $\sigma,\tau\in I_\pm$
 \be
 \bra \tau, 0|,  R^{L} |\sigma, y\ket= \bra \tau, 0|,  R^{L}T^{L} R^{L+3}  |\sigma, y\ket=0,
 \ee
we get for such vectors, 
\be \label{resolventexp}
\bra \tau, 0|,  R |\sigma, y\ket= \bra \tau, 0|,  R^{L}T^{L} R T^{L+3} R^{L+3} |\sigma, y\ket,
\ee
which is the geometric resolvent identity to be used below.

Define the boundary of $\hil^{\Lambda_L}$ by
\bea
\partial \hil_L & = & \{ (\alpha,\beta) \in (I_\pm\times\Z^d)^2\ | \
 \bra\alpha |, T^{L} |\beta\ket \not= 0\}  \\ \nonumber
& \subset & \{ (\alpha,\beta)\ | \  L-2 \le |\alpha_2| \le L+2, L-1 \le |\beta_2| \le L+1, |\alpha_2-\beta_2| \le 1\}.
\eea
Then, an expansion of (\ref{resolventexp}) over the boundaries of $\hil^{\Lambda_L}$ and $\hil^{\Lambda_{L+3}}$ gives
\bea
&&\bra \tau, 0|,  R |\sigma, y\ket = \\ \nonumber
&&\sum_{ (u,u') \in \partial \hil_L \atop (v,v') \in \partial \hil_{L+3} } \bra \tau, 0|,  R^{L}|u\ket\bra u|,T^{L} |u'\ket\bra u'|,R |v\ket\bra v|,T^{L+3} |v'\ket\bra v'|,R^{L+3} |\sigma, y\ket.
\eea
Taking the power $s<1$ and the expectation, we get
\bea \label{resolventexp2}
&&\E(|\bra \tau, 0|,  R |\sigma, y\ket|^s) \le c \, t^{2s}\\ \nonumber
&&\hspace{2cm}\times \sum_{(u,u') \in \partial \hil_L \atop (v,v') \in \partial \hil_{L+3}} \E \left( |\bra \tau, 0|,  R^{L}|u\ket|^s|\bra u'|,R |v\ket|^s |\bra v'|,R^{L+3} |\sigma, y\ket|^s \right).
\eea
Note that (\ref{dirres}) implies that only the vectors  $u\in \hil^{\Lambda_L}$ and $v' \in \hil^{\Lambda_{L+3}^C}$ are to be considered in the  sum above and give stochastically independent contributions. 
At this point we resort to a resampling argument to decouple the expectations and to the general estimate (\ref{thm31}) to get rid of the full resolvent term. The resampling argument requires $s\in (0,1/3)$, which we will assume from now on. The result is stated as Proposition 13.1 in \cite{HJS2} and it implies:

For every $s\in (0,1/3)$ there exists a constant $c <\infty$, depending on $s$, such that
\bea \label{eq:intext}
&&\E(|\bra \tau, 0|,  R |\sigma, y\ket|^s) \le c  \, t^{2s}  \\ \nonumber 
&&\hspace{1.cm}\times\sum_{|u\ket\in\hil^{\Lambda_L} \atop L-2\le |u_2|\le L+2} \E( |\bra \tau, 0|,  R^{L}|u\ket|^s)\sum_{|v'\ket\in\hil^{\Lambda_{L+3}^C} \atop L+2\le |v'_2|\le L+4} \E( |\bra v'|,R^{L+3} |\sigma, y\ket|^s )
\eea
uniformly in $z\not\in\U$ with $1/2<|z|< 2$, $L\in \N$ and $y\in \Z^d$ with $|y|\ge L+5$.\\

Next we relate $R^{L+3}$ to $R$ by means of 
\be
R^{L+3} = R + R^{L+3} T^{L+3} R,
\ee
and expand
\be
T^{L+3} = \sum_{(w,w')\in \partial \hil_{L+3}} |w\ket\bra w|, T^{L+3} |w'\ket\bra w'|.
\ee
Altogether we get
\bea \label{more}
&&\E( |\bra v'|,R^{L+3} |\sigma, y\ket|^s)  \le  \E( |\bra v'|,R |\sigma, y\ket|^s) \\ \nonumber 
&& \hspace{1.5cm} + c\,   t^s\,\sum_{(w,w')\in \partial \hil_{L+3}} \E( |\bra v'|,R^{L+3} |w \ket|^s  |\bra w'|,R |\sigma, y\ket|^s).
\eea

Then, another application of a resampling argument to factorize the expectations together with estimate (\ref{thm31}) eventually yields (\ref{itstep}), see Proposition 13.2 in \cite{HJS2}. \ep

\subsection{Iteration}
To obtain the sought for exponential estimate (\ref{expest}), we insert the estimate of Proposition \ref{finitevolresest} into that of Proposition \ref{ivsf} to get
\bea
\E(|\bra \tau, 0|,  R |\sigma, y\ket|^s) &\le& c (1+ct^{s}L^{(d-1)}) \frac{t^{2s}L^{2(d-1)}}{L^a}\max_{L+2\leq |x'_2|\leq L+4}\E(|\bra x'|,  R |\sigma, y\ket|^s)\nonumber \\
&\equiv&b(L)\max_{L+2\leq |x'_2|\leq L+4}\E(|\bra x'|,  R |\sigma, y\ket|^s),
\eea
with $t\leq c\|C-C_\pi\|=1/L^{2(ap+d)+a/s}$.
Taking $a>0$ large enough, with $0<s<1/3$ and $p>1/(1-s)$, we have
\be
t^{s}L^{(d-1)}
\ra 0 \ {\rm and } \ b(L)\ra 0 \ \  \mbox{if} \ L\ra\infty 
\ee
we can set $L_0\in \N^*$ large enough so that for $L= L_0$, we have $b(L_0)<1$ and 
\be\label{final}
\E(|\bra \tau, 0|,  R |\sigma, y\ket|^s) \le b(L_0) \max_{L_0+2\leq |x'_2|\leq L_0+4}\E(|\bra x'|,  R |\sigma, y\ket|^s).
\ee
Then it remains to invoke ergodicity in order to control the expectation at the right hand side by the same estimates as above, were the distance $|x'-y|$ can be decreased by $L$. Eventually, we join the points $0$ and $y$ in $\Z^d$ by a sequence of $n\leq c |y|/L_0$ boxes of side length $2L_0+1$ along which we iterate $n$ times the estimate
(\ref{final}) and use (\ref{thm31}) at the last step to eventually get
\be
\E(|\bra \tau, 0|,  R |\sigma, y\ket|^s) \le c b(L_0)^n\leq ce^{-\gamma |y|}, \ \mbox{with} \ \gamma=c|\ln(b(L_0))|/L_0 >0,
\ee
as requested. This finishes the proof of Theorem \ref{ame}, and that of Theorem \ref{dynloc}. \ep

\medskip

It is of interest to make the size of the perturbation parameter $\delta\geq \|C-C_\pi\|$ as large as possible in term of the scale $L$, by optimizing on the set of available parameters $a, s, p$, given $d$. Let us explain what we mean by this. The value $L_0>1$ is determined by the condition $L\geq L_0$ implies 
\be
b(L)\leq b_0/L^{\beta} \leq b_0/L_0^{\beta}= \eps<1, \ {\rm where } \ \beta=\left\{\begin{matrix}\beta_1 & {\rm if\  }d(1-2s)<a(1+2ps)+1\cr 
\beta_2 & {\rm otherwise,}\end{matrix}\right.
\ee
with
\bea
\beta_1&=& a(3+4ps)+2(1+d(2s-1)) \\ \nonumber \beta_2 &=&a(4+6ps)+3(1+d(2s-1))
\eea
and $b_0>1$.
The size of the perturbation is then determined by
\be
\delta(L_0)=1/L_0^{2(ap+d)+a/s}\equiv 1/L_0^{\rho}= (\eps/b_0)^{\rho/\beta}.
\ee
To maximize $\delta(L_0)$, we need to minimize ${\rho/\beta}$ over the parameters
$a, s, p$.
Without performing a complete analysis of the exponent ${\rho/\beta}$, we can obtain a value, uniform in the dimension $d$, by sending $a$ to infinity and then taking $p=(1+p')/(1-s)$, $s=(1-s')/3$, with $p', s'>0$ arbitrary small
\be
\rho/\beta=6/5+o(1).
\ee
\ep

{\bf Acknowledgements}
I wish to thank O. Bourget for useful discussions and the Mathematics Department of PUC Santiago, where part of this work was done, for its hospitality.

\end{document}